\documentclass[a4paper,11pt]{article}
\usepackage{pos}
\usepackage{graphicx}
\usepackage{subcaption}
\usepackage{siunitx}

\title{Status and prospects of the CORSIKA~8 air shower simulation framework}
\ShortTitle{Status of CORSIKA~8}

\manuallySeparateAuthors
\author*{Alexander Sandrock}
\author[\dag]{ for the CORSIKA~8 collaboration}

\affiliation{Bergische Universität Wuppertal,\\
  Gaußstraße 20, 42119 Wuppertal, Germany}
\notes{\note{The full author list can be found at \url{https://tinyurl.com/corsika8-202210}}}

\emailAdd{asandrock@uni-wuppertal.de}

\abstract{
The Fortran-versions of the CORSIKA air shower simulation code have been at the
core of simulations for many astroparticle physics experiments for the last 30
years. Having grown over decades into an ever more complex software,
maintainability of CORSIKA has become increasingly difficult, though its
performance is still excellent. In 2018, therefore a complete rewrite of
CORSIKA has begun in modern modular C++. Today, CORSIKA~8 has reached important
milestones with a full-fledged implementation of both the hadronic and
electromagnetic cascades, the ability to simulate radio and Cherenkov-light
emission from air showers and an unprecedented flexibility to configure
simulation media and their geometries.

This presentation will discuss the current status of CORSIKA~8, highlight the
new possibilities already available, and future prospects of this new air
shower simulation framework.
}

\FullConference{%
  *** 27th European Cosmic Ray Symposium - ECRS ***\\
  *** 25-29 July 2022 ***\\
  *** Nijmegen, the Netherlands ***
}

\begin{document}
\maketitle
\section{Introduction}
The air shower simulation code CORSIKA was originally developed for the KASCADE
experiment in the 1980s and has become a common reference frame for the
community \cite{Heck}. Thanks to decades of careful maintenance and development,
CORSIKA is at the core of air shower simulations in many astroparticle physics
experiments today.

As a hand-optimized code, CORSIKA shows an excellent performance, however, it
is also subject to some limitations. Maintenance of a monolithic Fortran code,
with program options heavily intertwined in the source code, is becoming
increasingly difficult. In addition, the possibilities of parallelization of the
Fortran version of CORSIKA are limited: MPI parallelization is available, but no
multi-threading or GPU parallelization are possible.

\begin{figure}
  \includegraphics[width=\textwidth]{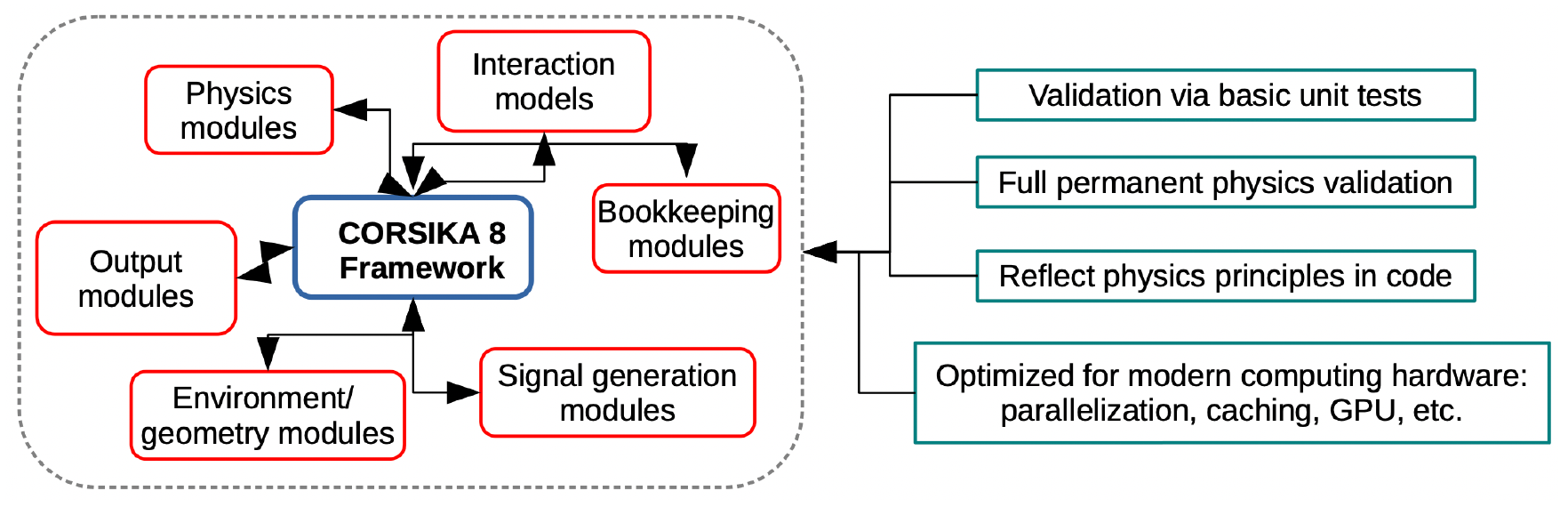}
  \caption{Overall structure of the CORSIKA~8 air shower simulation framework.
    From \cite{AugustoICRC}.}
  \label{fig:c8_framework}
\end{figure}

To address these issues, in 2018 a rewrite of CORSIKA in modern C++
(currently C++17) has begun, focussing on modularity and the needs and
possibilities of modern supercomputing environments
\cite{Reininghaus2019CORSIKA8,whitepaper}.
This effort is coordinated by KIT, but has a
strong community integration. The overall structure of this new version, called
CORSIKA~8, is shown in Figure~\ref{fig:c8_framework}.

\section{Status of CORSIKA~8}
Both hadronic and electromagnetic cascades are available, so CORSIKA~8 is now
capable of simulating complete air showers. Extensive validation by comparison
to CORSIKA~7 and other codes accompanies the development. Currently, CORSIKA~8
offers most of the possibilites of previous CORSIKA versions and already has
several capabilities, which were not available in the Fortran version, such as
full genealogy of particles, cross-media showers, and more flexible medium
definitions.

\subsection{Hadronic cascades}
\begin{figure}
  \includegraphics[width=\textwidth]{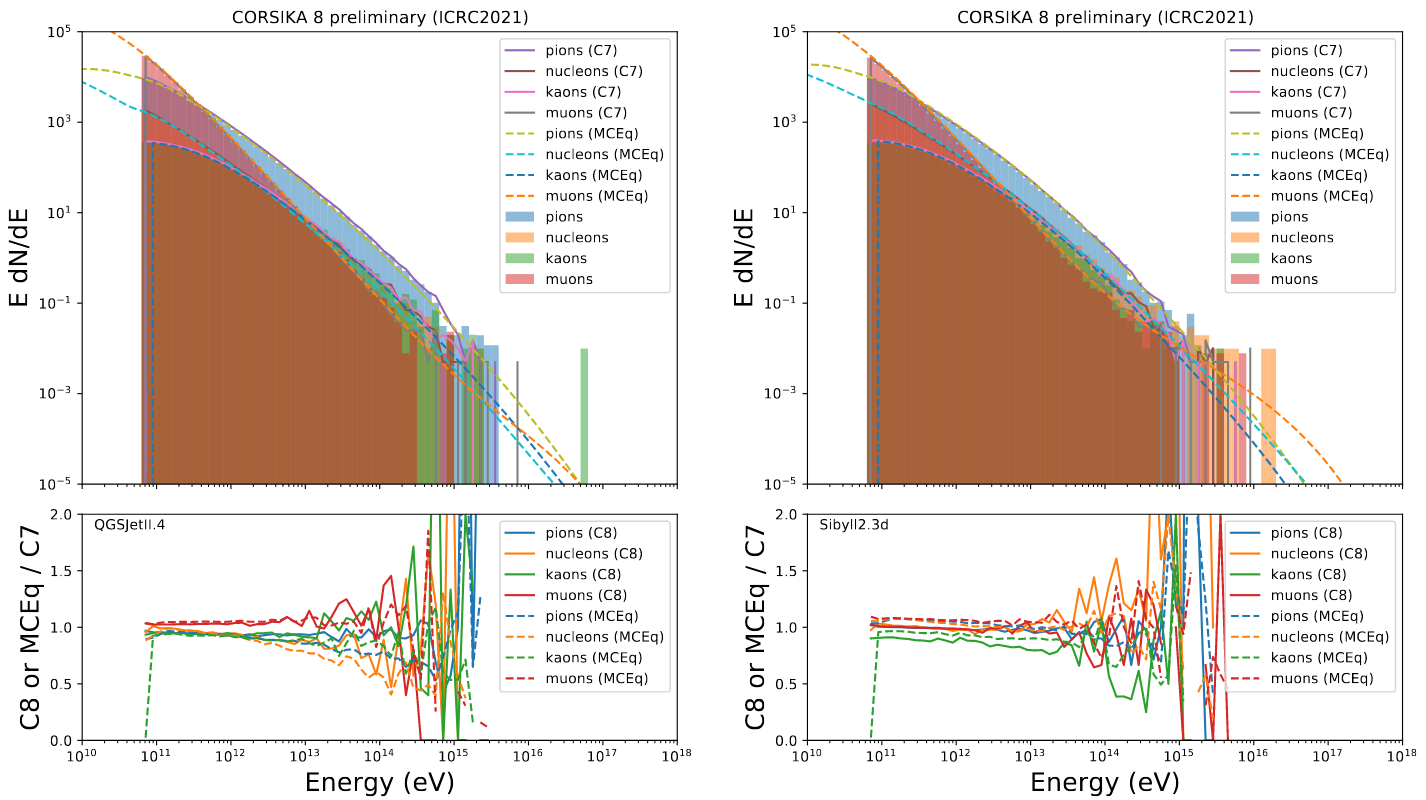}
  \caption{Comparison of particle spectra at observation level (\SI{1400}{m}
    a.\,s.\,l.) in a vertical proton shower at \SI{e18}{eV} between CORSIKA~8,
    CORSIKA~7, and MCEq.
    From \cite{UlrichICRC}.}
  \label{fig:ulrich_fig2}
\end{figure}

Currently, the available hadronic interaction models in CORSIKA~8 are
QGSjet-II-04, EPOS-LHC, and Sibyll~2.3d at high energies, UrQMD at low energies,
and decays are treated either with Sibyll~2.3d or PYTHIA~8. Comparisons of
particle spectra between CORSIKA~8, CORSIKA~7, and MCEq have been presented in
\cite{UlrichICRC}, and show a good agreement between the
different codes (cf. Figure~\ref{fig:ulrich_fig2}).

\begin{figure}
  \includegraphics[width=\textwidth]{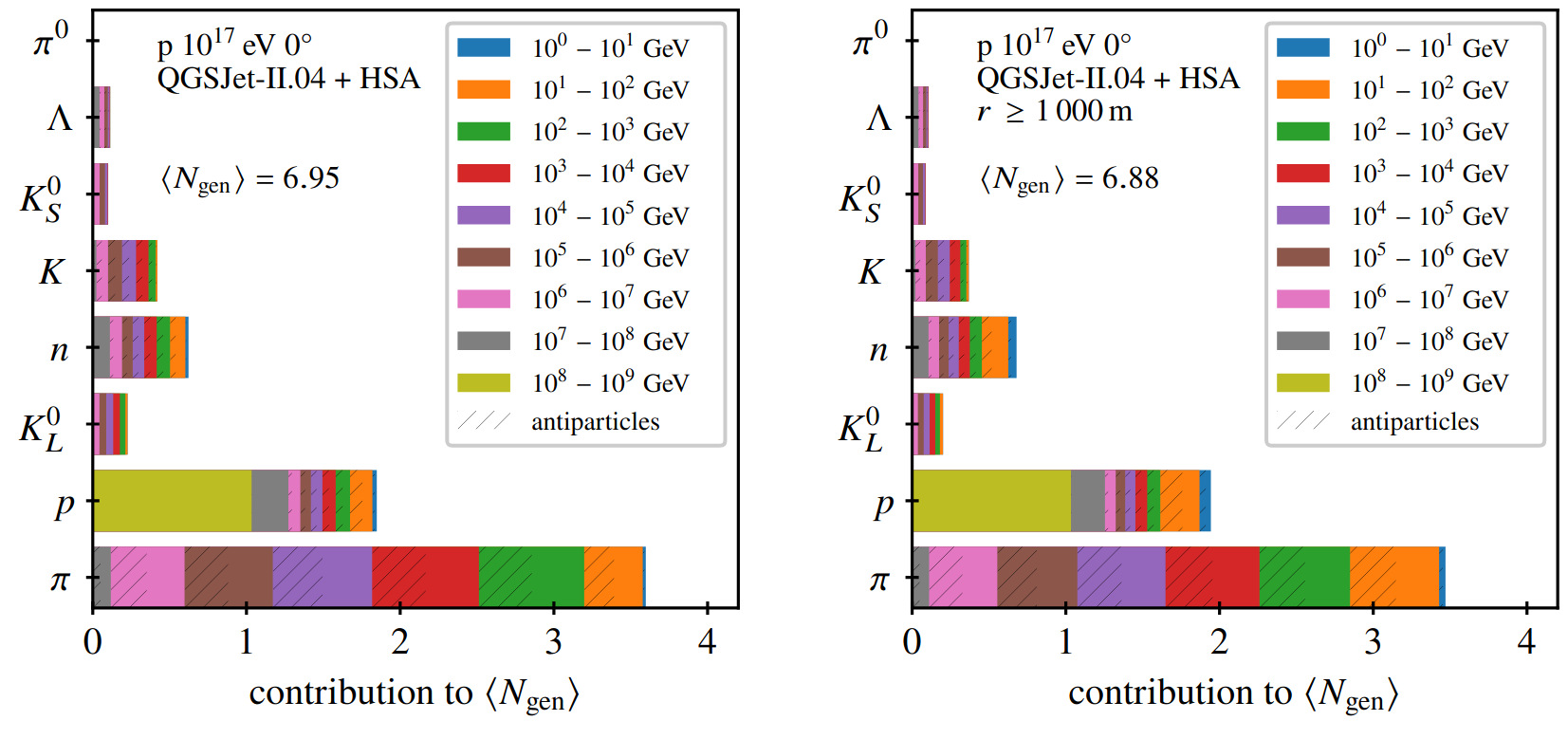}
  \caption{Muon ancestor particle distributions by species and energy.
    From \cite{ReininghausICRC}.}
  \label{fig:reininghaus_fig3}
\end{figure}

A new possibility of CORSIKA~8 is the genealogy of particles. While current
air shower simulation codes have only the possibility to identify the mother
and grandmother particle, in CORSIKA~8 the complete genealogy of particles is
available (cf. Figure~\ref{fig:reininghaus_fig3}). A detailed report on muon
genealogy has been published in \cite{ReininghausICRC}.

\begin{figure}
  \begin{center}
    \includegraphics[width=0.7\textwidth]{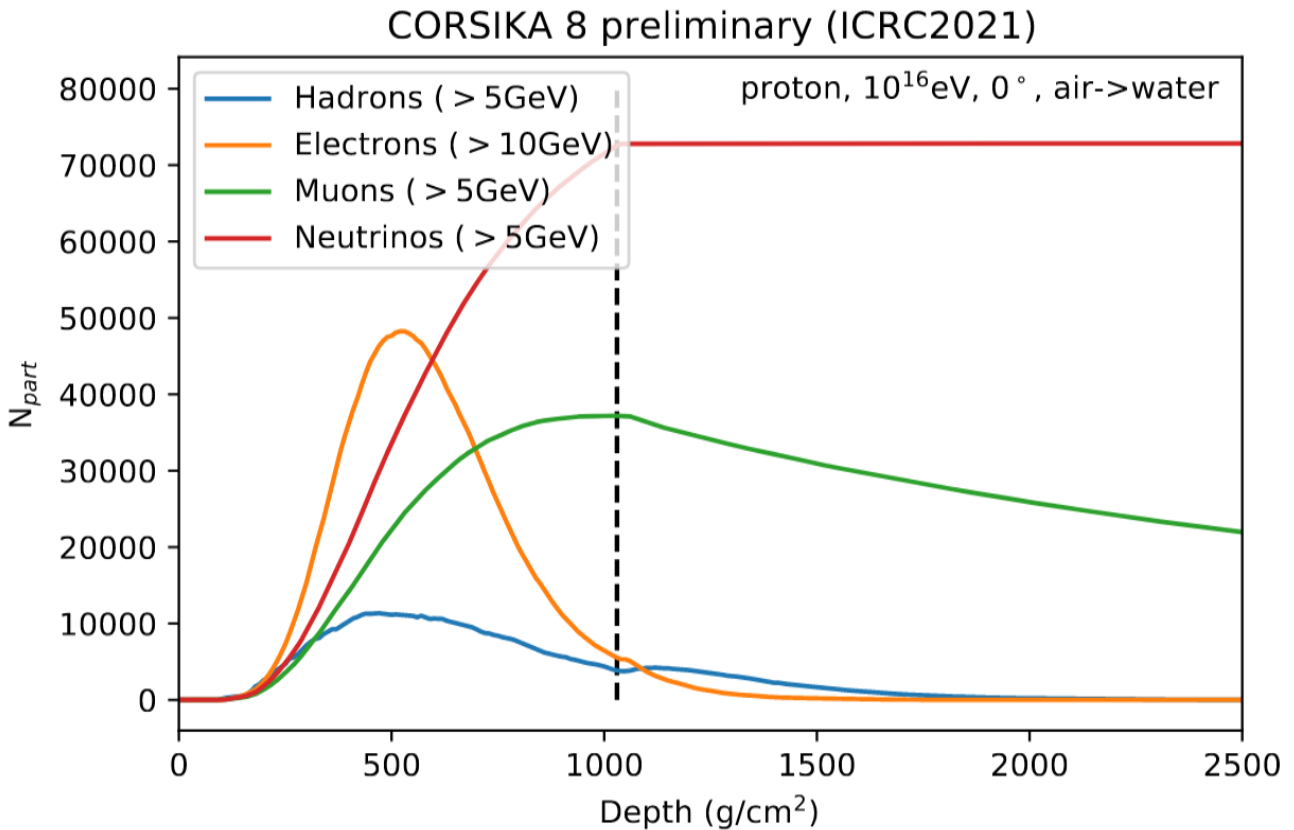}
  \end{center}
  \caption{Vertical \SI{e16}{eV} proton shower transitioning from air to water.
    From \cite{UlrichICRC}.}
  \label{fig:ulrich_fig6b}
\end{figure}

Another novelty is the possibility to consistently treat cross-media showers,
e.\,g. cosmic ray showers transitioning from air to water or ice, inside a
common framework. As an example, the longitudinal profile of a vertical
\SI{e16}{eV} proton shower transitioning from air to water is shown in
Figure~\ref{fig:ulrich_fig6b}.

\subsection{Electromagnetic cascades}
In CORSIKA~7, electromagnetic cascades are treated using a modified version of
EGS~4 \cite{EGS4}, a Mortran code that was deeply integrated into the CORSIKA
source code; beside the processes contained in the original EGS~4, the muon pair
production process $\gamma \rightarrow \mu\mu$, the photohadronic interaction
$\gamma N \rightarrow X$, and (optionally) the Landau-Pomeranchuk-Migdal effect
have been added. Alternatively, the analytic NKG treatment of electromagnetic
cascades is available.

\begin{figure}
  \includegraphics[width=\textwidth]{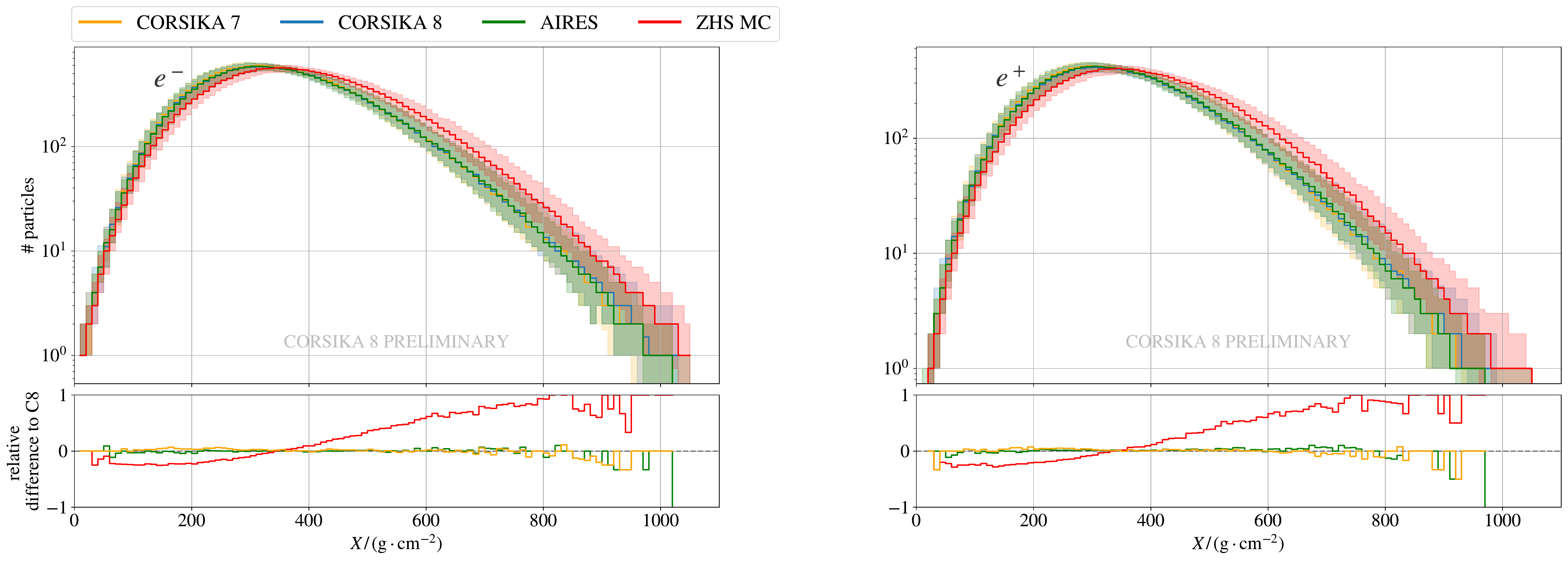}
  \caption{Longitudinal profile of \SI{1}{TeV} electromagnetic showers
    in C7, C8, AIRES, and ZHS MC. From \cite{AlameddineICRC}.}
  \label{fig:em_longitudinal}
\end{figure}

\begin{figure}
  \begin{center}
    \includegraphics[width=0.7\textwidth]{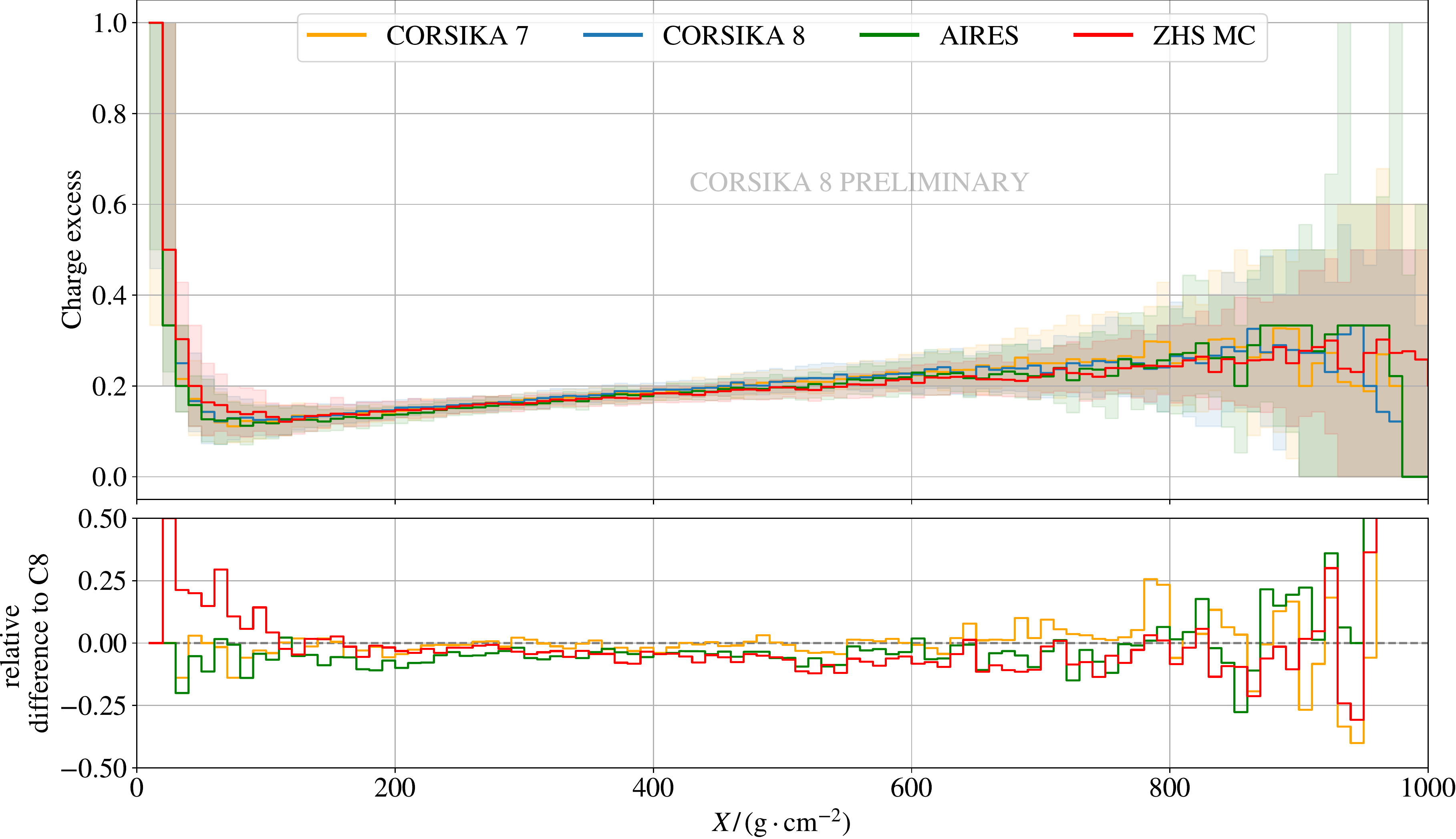}
  \end{center}
  \caption{Charge excess of \SI{1}{TeV} electromagnetic showers
    in C7, C8, AIRES, and ZHS MC. From \cite{AlameddineICRC}.}
  \label{fig:em_charge_excess}
\end{figure}

In CORSIKA~8, electromagnetic cascades are simulated using the lepton propagator
PROPOSAL \cite{Koehne2013PROPOSAL,Dunsch2019PROPOSAL,Alameddine2020ICPPA}, a
modular C++14 library with Python bindings, which can propagate electrons,
positrons, photons as well as muons and tau-leptons. The LPM effect is currently
 available only
in media with homogeneous density; for inhomogeneous media, this is currently
under development. The rather good agreement between CORSIKA~7, CORSIKA~8,
 AIRES,
and the ZHS air shower Monte Carlo code is shown for the longitudinal profile and the charge excess of
\SI{1}{TeV} electromagnetic showers in Figures~\ref{fig:em_longitudinal} and
\ref{fig:em_charge_excess}.

\subsection{Radio and Cherenkov emission}
The aim of the CORSIKA~8 radio module is to overcome limitations of previous
CORSIKA versions, in particular to be able to simulate 
 the reflected radio signal of downwards-going showers, to
consider the ray curvature in the atmosphere, and to simulate showers crossing
from air to dense media.

\begin{figure}
  \begin{center}
    \includegraphics[width=0.7\textwidth]{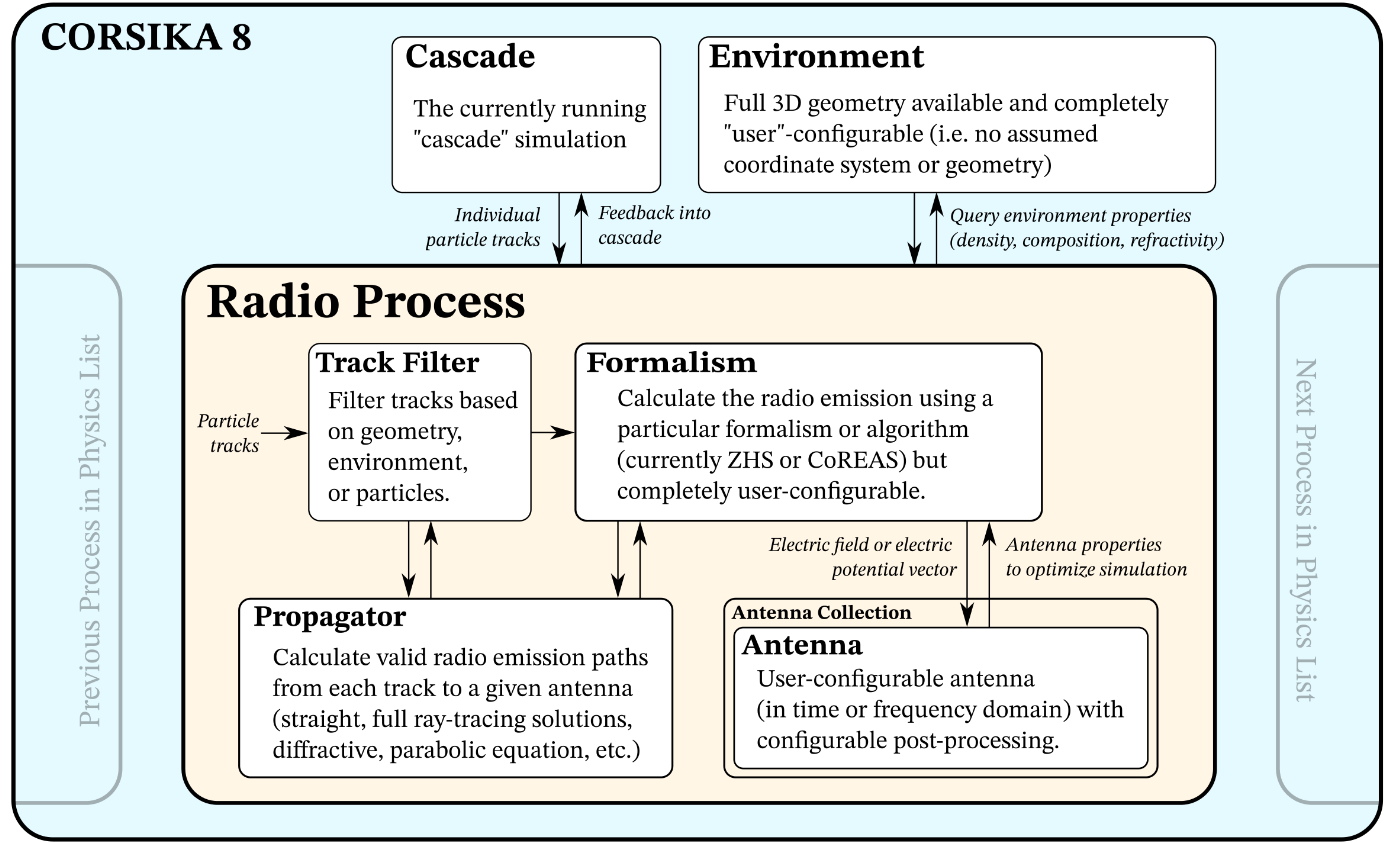}
  \end{center}
  \caption{Schema of radio emission calculation in CORSIKA~8.
    From \cite{KarastathisICRC}.}
  \label{fig:radio_schema}
\end{figure}

\begin{figure}
  \begin{subfigure}{0.45\textwidth}
    \includegraphics[width=\textwidth]{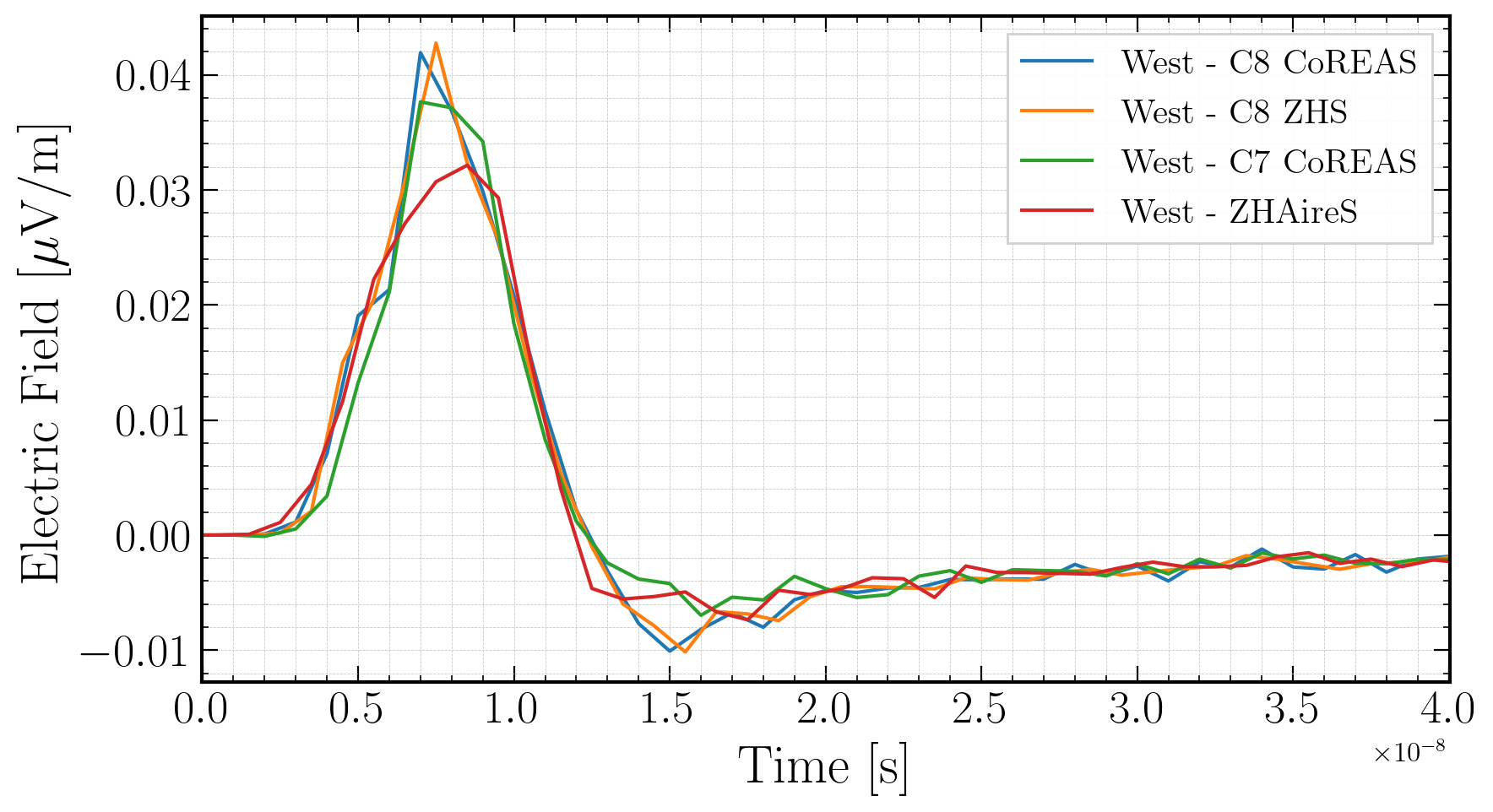}
  \end{subfigure}
  \hfill
  \begin{subfigure}{0.45\textwidth}
    \includegraphics[width=\textwidth]{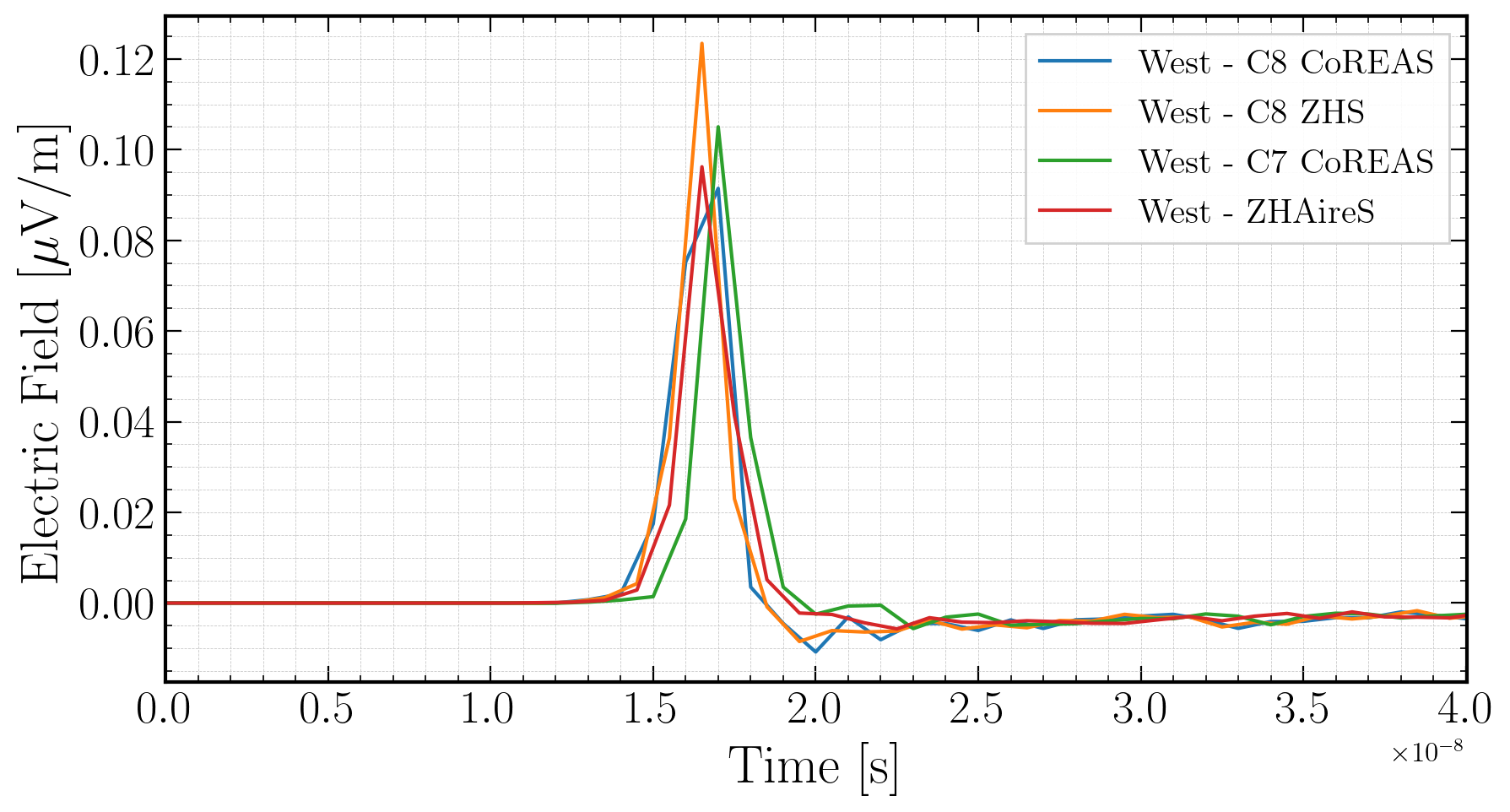}
  \end{subfigure}
  \caption{Radio pulses from a \SI{10}{TeV} electromagnetic shower
    at \SI{50}{m} (left) and \SI{200}{m} (right) distance from shower core,
    west polarization. From \cite{KarastathisArena}.
    }
  \label{fig:radio_pulse}
\end{figure}

\begin{figure}
  \begin{subfigure}{0.45\textwidth}
    \includegraphics[width=\textwidth]{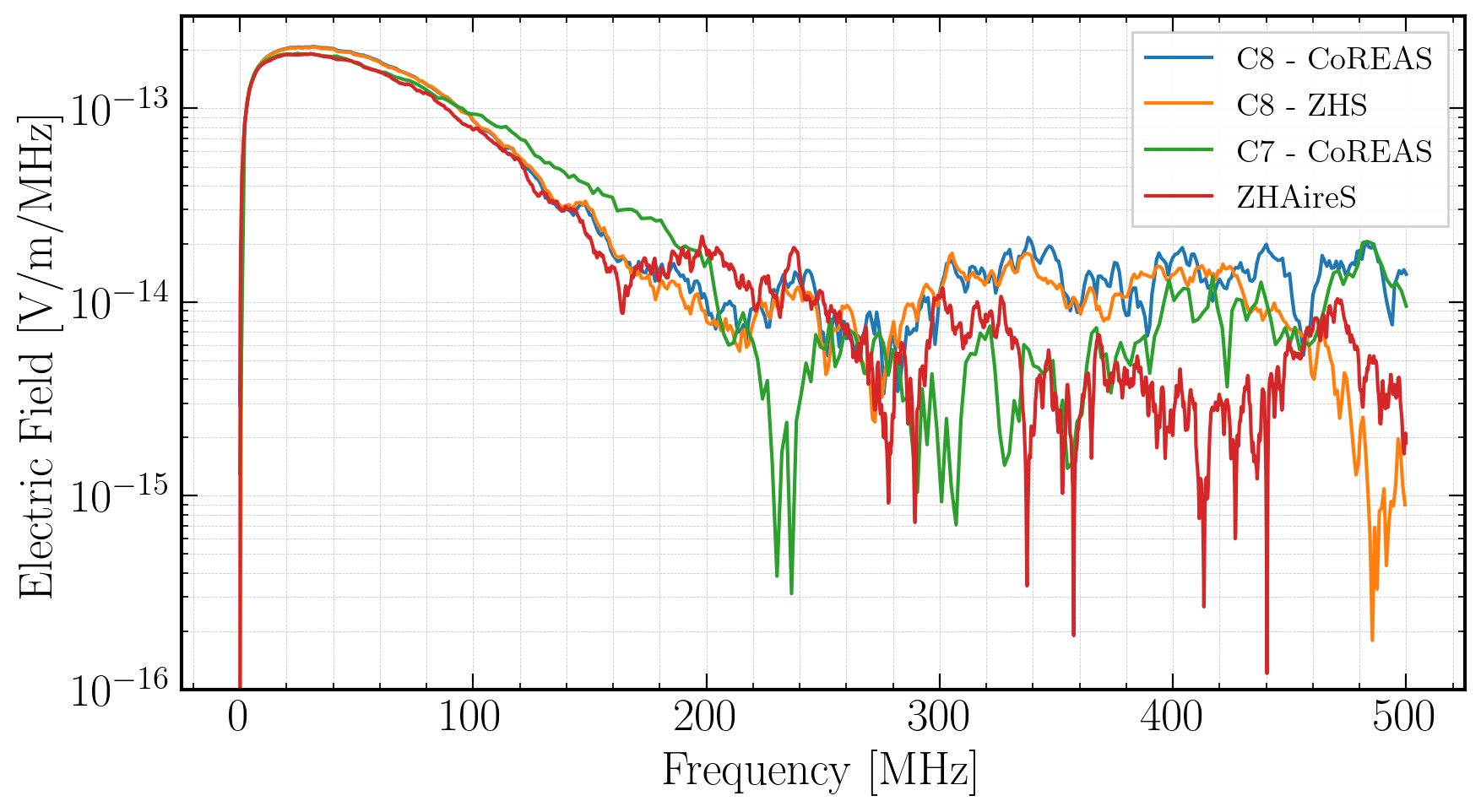}
  \end{subfigure}
  \hfill
  \begin{subfigure}{0.45\textwidth}
    \includegraphics[width=\textwidth]{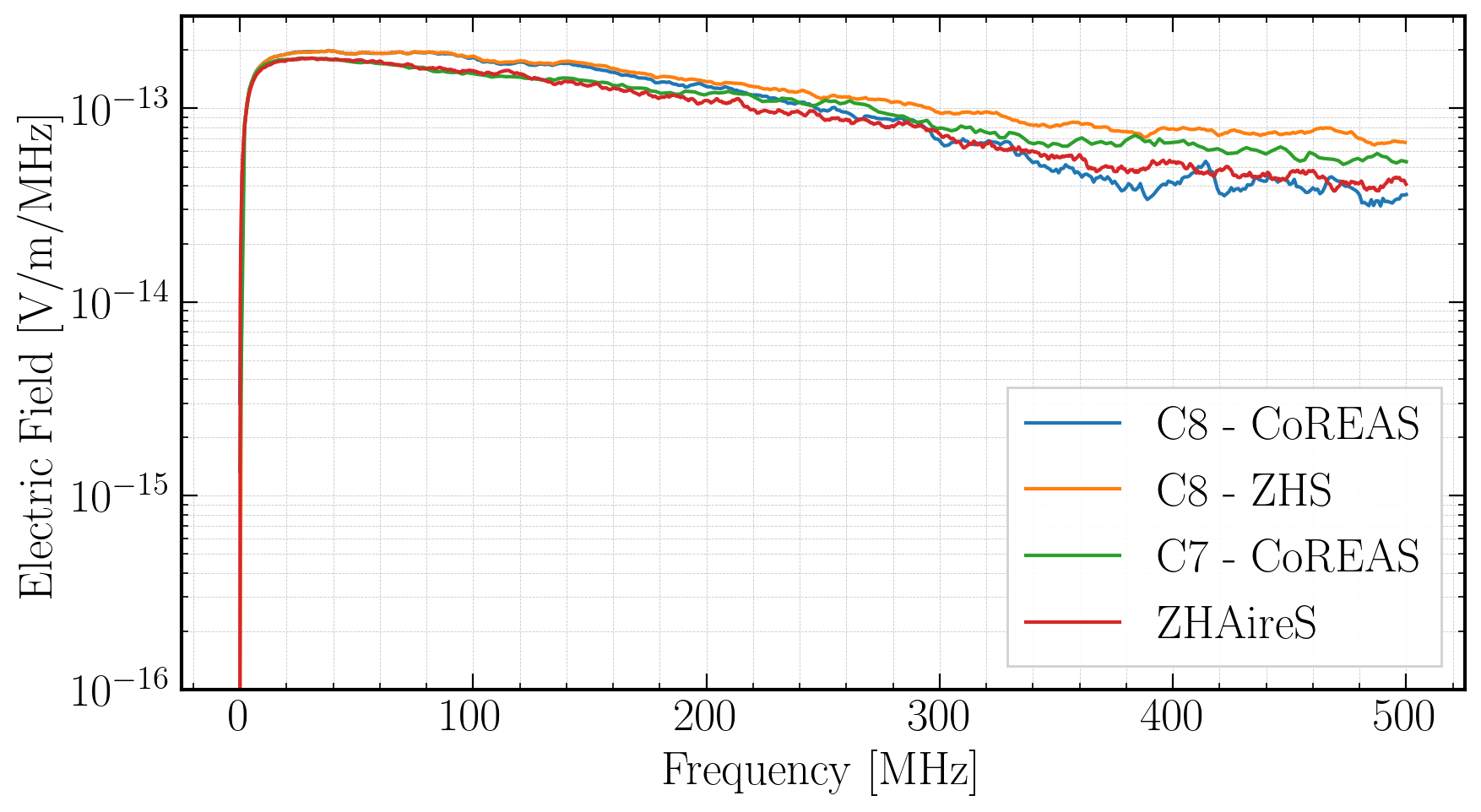}
  \end{subfigure}
  \caption{Radio frequency spectra of a \SI{10}{TeV} electromagnetic shower
    at \SI{50}{m} (left) and \SI{200}{m} (right) distance from shower core,
    west polarization. From \cite{KarastathisArena}.
    }
  \label{fig:radio_spectrum}
\end{figure}

To calculate the radio emission, two algorithms have been implemented, the
CoREAS algorithm as in CORSIKA~7, and the ZHS algorithm as in ZHAireS; the radio
emission calculated according to both formalisms is in good agreement. The radio
emission is fully implemented as a process, with particle filtering, the used formalism,
the radio wave propagator, and the antenna configuration configurable by the
user. A schematic of the radio emission calculation is shown in Figure~%
\ref{fig:radio_schema}.
A detailed account of the radio implementation in
CORSIKA~8 was given in \cite{KarastathisICRC}.
The agreement of the radio emission between CORSIKA~7, ZHAireS and the two
formalisms in CORSIKA~8 is illustrated in Figures~\ref{fig:radio_pulse} and
\ref{fig:radio_spectrum}, which provide update results to \cite{KarastathisICRC}. 


\begin{figure}
  \begin{center}
    \includegraphics[width=0.7\textwidth]{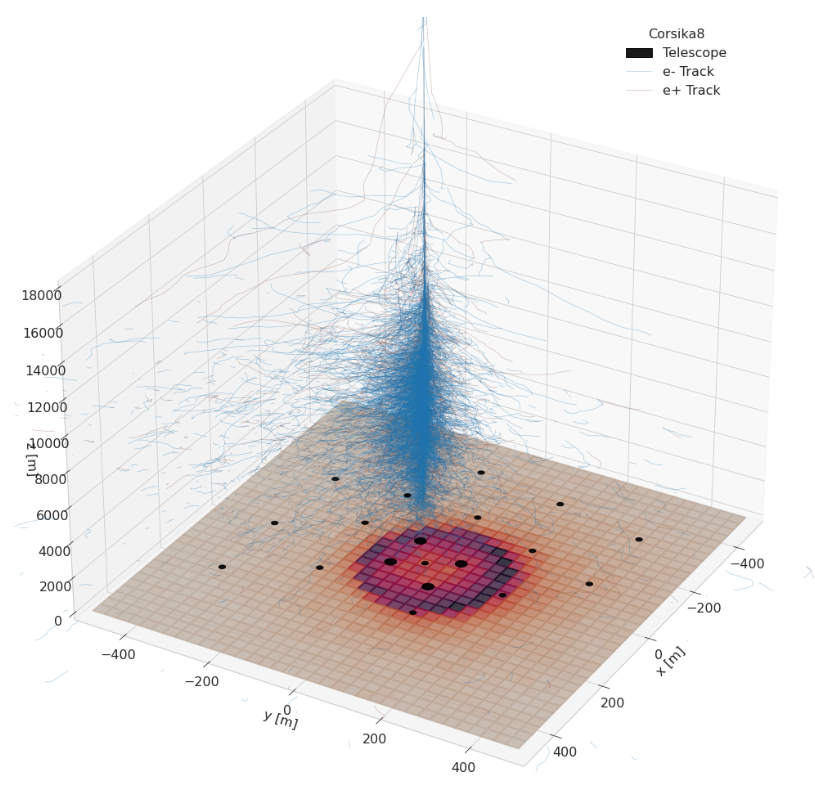}
  \end{center}
  \caption{\SI{1}{TeV} shower with ground level distribution of
    Cherenkov light. From \cite{BaackICRC}.}
  \label{fig:cherenkov}
\end{figure}

The Cherenkov module of CORSIKA~8 provides two implementations for the
calculation of Cherenkov emission. The calculations of these are in good
agreement with each other and CORSIKA~7. One of the implementations is
vectorized, the other uses GPU parallelization. The ground level distribution
of Cherenkov light from a \SI{1}{TeV} shower is shown in Figure~%
\ref{fig:cherenkov}. A detailed account of the Cherenkov modules was given in
\cite{BaackICRC,CarrereCHEP}.

\section{Conclusion}
CORSIKA~8 is an open source project, with the source code available on the
KIT gitlab server\footnote{%
\url{https://gitlab.iap.kit.edu/AirShowerPhysics/corsika/}}.
Current directions of development for CORSIKA~8 are the performance optimization
of the code, an improved treatment of multiple scattering, the implementation of
the Landau-Pomeranchuk-Migdal effect in inhomogeneous media, the development of
interfaces to PYTHIA~8\footnote{First results with the interface to PYTHIA~8
were presented after the symposium at UHECR2022 \cite{ReininghausUHECR2022}.},
FLUKA, and SOPHIA, and the implementation of photohadronic
interactions at low energies.

CORSIKA~8 is now capable of simulating all components of extensive air showers,
provides most possibilities of CORSIKA~7 and already has features going beyond
earlier versions. Now is a great time for developers to join; the first release
for end users is tentatively planned for mid-2023.

\section*{Acknowledgments}
The speaker acknowledges funding by the Deutsche Forschungsgemeinschaft (DFG) --
Project number SA~3867/2-1.

\bibliographystyle{JHEP}
\bibliography{references.bib}
\end{document}